\documentclass[twocolumn]{aastex62}
\usepackage{newtxtext,newtxmath}
\usepackage{natbib, amsmath, color}
\usepackage{graphicx, hyperref,lastpage}

\usepackage[T1]{fontenc}

\DeclareRobustCommand{\VAN}[3]{#2}
\let\VANthebibliography\thebibliography
\def\thebibliography{\DeclareRobustCommand{\VAN}[3]{##3}\VANthebibliography}

\usepackage{graphicx}	
\usepackage{amsmath}	
\usepackage{amssymb}	
\usepackage{xcolor,ulem}

\shorttitle{Merger's Viewing Angle}
\shortauthors{Nakar \& Piran}

\begin{document}
\title[Merger's Viewing Angle]{ Afterglow constraints on the viewing angle of binary neutron star mergers and determination of the Hubble constant}
\author{Ehud Nakar}
\affil{School of Physics \& Astronomy \\ Tel Aviv University, Tel Aviv 69978, Israel
}
\author{Tsvi Piran}
\affiliation{Racah Institute of Physics \\
The Hebrew University of
			Jerusalem, Jerusalem 91904, Israel}





\begin{abstract}
One of the key properties of any binary is its viewing angle (i.e., inclination), $\theta_{\rm obs}$. In binary neutron star (BNS) mergers it is of special importance due to the role that it plays in the measurement of the Hubble constant, $H_0$.  The   opening angle of the jet that these mergers   launch, $\theta_j$, is also of special interest. Following the detection of the first BNS merger, GW170817, there were numerous attempts to estimate these angles using the afterglow light curve, finding a wide range of values. 
Here we provide a simple formula for the ratio $\theta_{\rm obs}/\theta_j$ based on the afterglow light curve and  show that this is the only quantity that can be determined from the light curve alone.
Our result  explains the inconsistency of the values found by the various studies of GW170817 that were largely  driven by the different priors taken in each study. Among the additional information that can be used to  estimate  $\theta_{\rm obs}$ and $\theta_j$, the most useful is a VLBI measurement of the afterglow image superluminal motion. An alternative is an identification of the afterglow transition to the sub-relativistic phase. These observations are possible only for mergers observed at small viewing angles,   whose afterglow is significantly brighter than the detector's threshold. 
We discuss the implications of these results to measurements of $H_0$  using GW observations.  We show that while the viewing angle will be measured only in a small fraction of future BNS mergers, it can significantly reduce the uncertainty in $H_0$ in each one of these events, possibly to a level of 4-5\%. 
\end{abstract}

\vskip 0.5cm



\section{Introduction}
The combination of gravitational waves (GW) and electromagnetic (EM) emission from Binary Neutron Star (BNS) mergers offers unique opportunities (for a review see \citealt{nakar2019}). The multi-messenger information enables us to explore new physics that could not have been explored before. Some of this new information depends on accurate knowledge of the system geometry, including in particular the viewing angle at which we observe it. Here we discuss pitfalls in the geometry determination and their remedies. 

Already in the eighties \cite{Schutz1986} suggested to use BNS mergers as standard gravitational wave sirens and obtain an independent measurement of the Hubble constant, $H_0$. Such a measurement is of great interest today in view of the tension \citep[e.g.][]{Feeney2018} between local \citep{Riess2016}  and distant \citep{Planck2016} estimates of $H_0$.  This potential was demonstrated beautifully with GW170817, the first detection of gravitational waves from a BNS merger,  with which the LVC team estimated $H_0=74^{+16}_{-8}$ km/s/Mpc \citep{Abbott2017hubble}.  The degeneracy in the gravitational waves signal between the viewing angle and the source distance is the  dominant source of uncertainty in this measurement. Hence, a precise determination of the viewing angle  is of great importance. Indeed, 
\cite{Hotokezaka2019} obtained an improved estimate of  $H_0=70.3^{+5.3}_{-5}$ km/s/Mpc using  a measure of the viewing angle of GW 170817 based on VLBI observations of the afterglow radio image \citep{Mooley+18b}.
While still insufficient to resolve the above mentioned tension, this result demonstrates both the potential of this method and the importance of the exact determination of the viewing angle. 

The  system geometry, which in addition to the viewing angle includes the angular structure of the jet and the ejecta, would also be of extreme importance for understanding the detailed physics of the merger. This in turn can shed light on questions ranging from the equation of state of high density matter, via the sites of r-process nucleosynthesis to the mechanisms of jet acceleration and the link to short gamma-ray bursts. 

In view of  its importance, there have been numerous attempts to estimate the system geometry of  GW170817 based on the available  data. Some information concerning the viewing angle, $\theta_{\rm obs}$, is available directly from the GW signal if the value of $H_0$ is assumed \citep{Abbott170817Detection}. But these limits are not sufficiently accurate and clearly  are not useful when one wants to  determine $H_0$. 
The viewing angle can also be determined using the polarization of the GW signal. However,  the amplitude ratio between the two polarizations is sensitive enough to be measured only when $\theta_{\rm obs} \gtrsim 70^o $ \citep{Chen2018}. 

Almost all the efforts to reveal the geometry of GW170817 have been  based on model fitting to the  EM afterglow light curve\footnote{\cite{Dhawan+2020}  suggested a method to estimate the viewing angle using the macronova/kilonova signal. However,
while the afterglow light curve is based on rather clear physics the macronova/kilonova light curve involve uncertainties in all aspects, ranging from the composition of the ejecta to its three dimensional distribution, unknown opacities of the relevant heavy elements and complicated radiation transfer calculations  (see \citealt{nakar2019} for a discussion of the various uncertainties). Thus, a concern arises that due to these numerous uncertainties it might be extremely difficult to control the systematics that may arise using this method.}
\citep{Margutti+17,Troja+17,Alexander+17,Haggard+17,Lazzati+2018,Alexander+18,Margutti+18,Dobie+18,DAvanzo+2018,Troja+18,Mooley+18a,Granot+18,Gill+2018,Lyman2018,Lamb+2019,Fong+19,Troja+19,Wu2019,Hajela+19,Takahashi2020,Ryan2020}. All recent studies agree that GW170817 have launched a relativistic jet that broke out of the ejecta successfully and then interacted with the circum-merger medium to produce the afterglow. They all find that the jet had an angular structure with a narrow highly energetic core, with a jet opening $\theta_j$, which is observed from a viewing angle $\theta_{\rm obs} \gg \theta_j$.
Each of these studies obtained a different estimate of $\theta_{\rm obs}$, and  $\theta_j$   (see \S \ref{sec:GW170817} and Figure \ref{fig:fitting}).  The viewing angles found by different studies are in the range of   $14^o$ to $38^o$ and the jet opening angle is in the range of  $2.5^o$ to $8^o$. The typical errors quoted in these studies are a few degrees on the viewing angle and a fraction of a degree on the jet opening angle. Thus, the various estimates are often highly inconsistent with each other.
An exception is the work of \cite{Mooley+18b} who used the motion of the radio image as measured in VLBI observations  \citep[see also][]{Ghirlanda+19,Hotokezaka2019} in addition to the light curve to constrain the system geometry. 

The immediate question that arises is - how analyses of the same data set with similar modeling methods provide different and inconsistent results.
More importantly, it brings the question how reliable are measurements of the viewing angle that are based on GW170817-like afterglow light curves alone. Here we address these questions showing that the light curve by itself is insufficient to obtain a useful measure on  $\theta_{\rm obs}$ and $\theta_j$ separately and it can only constrain their ratio, $\theta_j/\theta_{\rm obs}$. We show that the main reason is that as long as the jet is relativistic there is an intrinsic degeneracy in the shape of the light curve between $\theta_j, \theta_{\rm obs}$ and the Lorentz factor of the emitting region, $\Gamma$.  A proper scaling of the three leads to similar light curves. Since the Lorentz factor is determined by the unknown ratio of the jet energy and the circum-burst density, different geometries can generate similar light curves by varying the value of this ratio. This degeneracy is similar to the degeneracies in the light curve discussed by \cite{Beniamini2020}.

{The correlation between $\theta_{\rm obs}$ and $\theta_j$ that is found in light curve modeling was already identified by \citet{Ryan2020} and \citet{Takahashi2020}. Both papers focus on the information carried by the rising phase of the light curve (denoted as the ``structured phase'' by \citealt{Ryan2020}), but they also carry out a fit to the entire light curve of GW170817. Both papers find that there is some level of degeneracy between $\theta_{\rm obs}$ and $\theta_j$ and that in order to fit the light curve of GW170817 a larger viewing angle requires a larger jet angle. Moreover, \citet{Ryan2020} find that they are able to constrain the ratio $\theta_j/\theta_{\rm obs}$ much better than each of the angles separately. They conclude that in order to robustly measure $\theta_{\rm obs}$ from the afterglow light curve one must be sure to allow for a large variety of jet structure profiles during the fitting process. Here we show that constraining the viewing angle is even more difficult. The degeneracy of light curves such as that of GW170817 afterglow is complete and any attempt to constrain the viewing angle based on the light curve alone is futile.  The light curve alone cannot provide any useful information on $\theta_{\rm obs}$ or $\theta_j$ apart for their ratio and the fact that $\theta_{\rm obs} \lesssim 1$ rad. }

We show that additional information is needed in order to break this degeneracy and measure $\theta_{\rm obs}$ and $\theta_j$. The most useful option is VLBI observations that measure the superluminal motion of the centroid of the radio image. This, in turn, enables determination of  the Lorentz factor  at the time of the afterglow peak. Alternatively, late observations of the light curve transition to the sub-relativistic phase can also break this degeneracy. 

The structure of the paper is as follows. In order to demonstrate the aforementioned degeneracy we  begin in \S \ref{sec:numerical} with two specific examples. We use semi-analytic light curves of  Gaussian structured jets and top-hat jets with different opening angles and different viewing angles.
We stress that our goal is not to carry out another advanced numerical simulations. Our goal is to provide a general explanation for the nature of this degeneracy, that is found in all current simulations. We do so in \S \ref{sec:analytic} using general analytic  considerations,  and show that
the width of the peak provides a measure of  $\theta_j/\theta_{\rm obs}$. In this section we also summarise recent results on what can (and what cannot) be learned from the rising part of the light curve.
In \S \ref{sec:breaking} we discuss how a measurement of the superluminal motion of the centroid of the image or observation of the transition to the sub-relativistic phase can remove the degeneracy. We examine previous analysis of  GW170817 in \S \ref{sec:GW170817} showing that all previous analysis of this light curve exhibit this degeneracy, even though different models and different numerical schemes have been used. In \S \ref{sec:Hubble} we consider the implications of our results to $H_0$ measurements. We conclude, pointing out the broader implications of these uncertainties on our (lack of) ability to determine other jet parameters just from the light-curve and summarize in  \S \ref{sec:conclusions}. 

\section{Degeneracy of afterglow light curves - specific examples of a general characteristic} 
\label{sec:numerical}

\begin{figure}
\centering
\includegraphics[scale=.4]{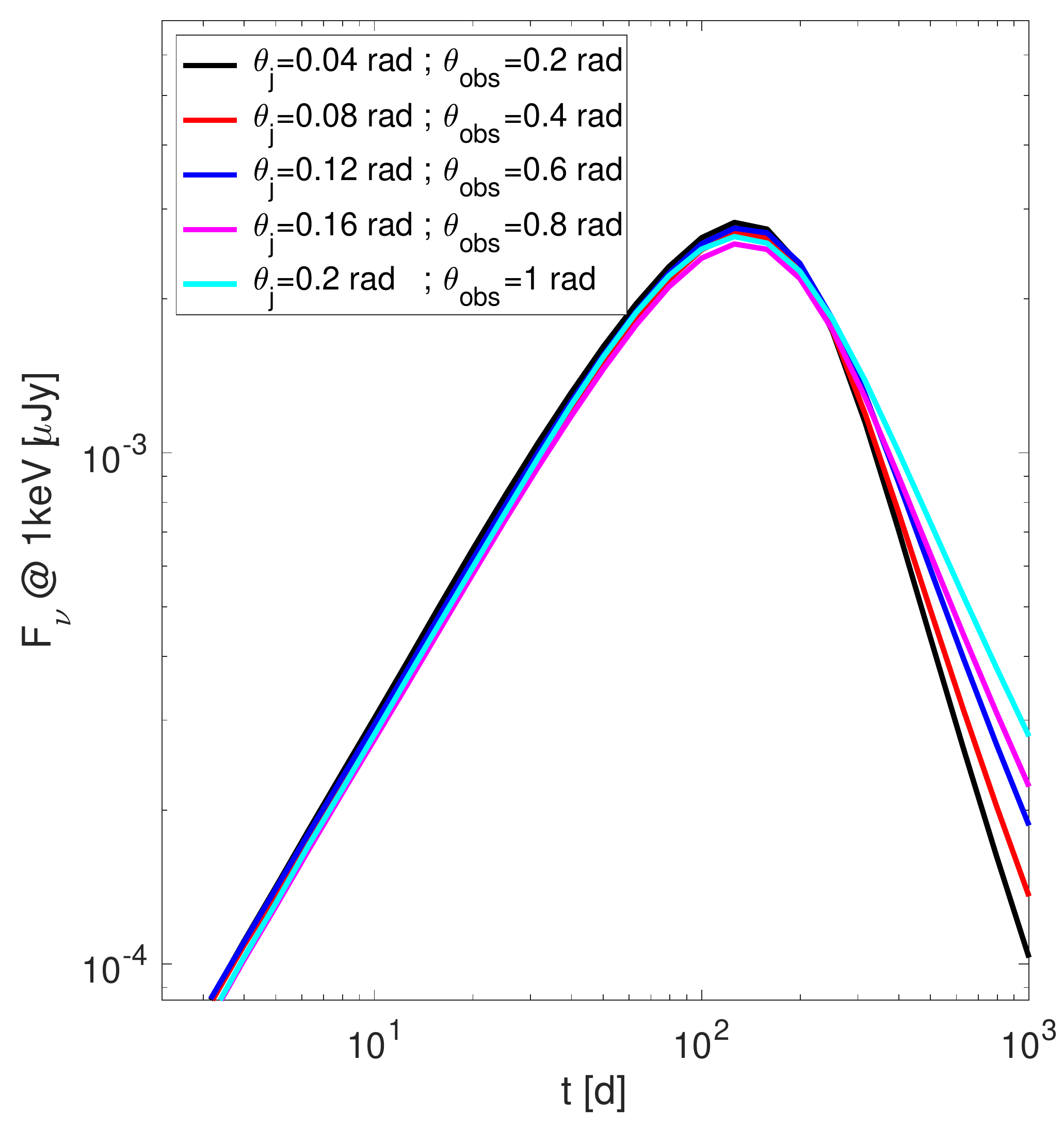}
\vskip -0.2cm
\caption{{ Afterglow light curves for jets that have similar structure but different opening angles and viewing angles, such that the ratio $\theta_{\rm obs}/\theta_j$ remains constant. All the curves are practically the same from the early rising phase and until long after the  peak. Curves start deviating only at late time, one after the other, when the blast wave that generates each curve approaches the Newtonian regime. The curves shown in this example are for Gaussian jets and the parameters were chosen so they all provide a reasonable fit to GW170817 afterglow  (see text).  In \S \ref{sec:analytic} we show that the degeneracy shown in this example is 
general and it arises for different jet structures, different afterglow models and different spectral ranges (as long as the entire curve is on the same power-law segment). }}
\vskip -0.4cm
\label{fig:Gaussian}
\end{figure}

{ We  begin by showing in Figure \ref{fig:Gaussian} a specific example that demonstrates a general characteristic of afterglow light curves. In this figure we present several afterglow light curves of the commonly used structured Gaussian jet,} with a slight modification, $E_{\rm iso}(\theta)= E_0 {\rm exp}[-(\theta/\theta_j)^{1.8}]$  where  $E_0$ and $\theta_j$ are constants and $\theta$ is the angle with respect to the jet symmetry axis\footnote{The slight modification from a Gaussian, a power of 1.8 instead of 2 in the exponent, was taken to obtain a slightly improved fit to the data of GW170817}. The jet opening angle in this model is roughly the jet core, $\theta_j$.
The light curve from this jet is calculated using standard afterglow theory  \citep[see][for a description of the code] {Soderberg2006}. This specific calculation assumes no significant lateral motion within the jet, so lateral expansion is ignored and along each direction the jet is assumed to propagate as if it is a part of a spherical blast wave. This approximation is used in many light curve calculations of structured jets \citep[e.g.,][]{Lazzati+2018,Lamb2018,Lyman2018}.  
{ The parameters in all curves were chosen so they all fit the afterglow of GW170817. All the parameters are within the range expected for BNS merger afterglows: $E_0=10^{52}$ erg, $p=2.1$  and $\epsilon_e=0.1$ while $n$ and $\epsilon_B$ vary between the different curves, so the time and flux at the peak of all the curves coincide. $n$ varies in  the range $3 \times 10^{-6}-0.7 {\rm~cm^{-3}}$ and $\epsilon_B$ varies in the range $10^{-5}-0.013$. The burst distance is 40 Mpc}

\begin{figure}
\centering
\includegraphics[scale=.4]{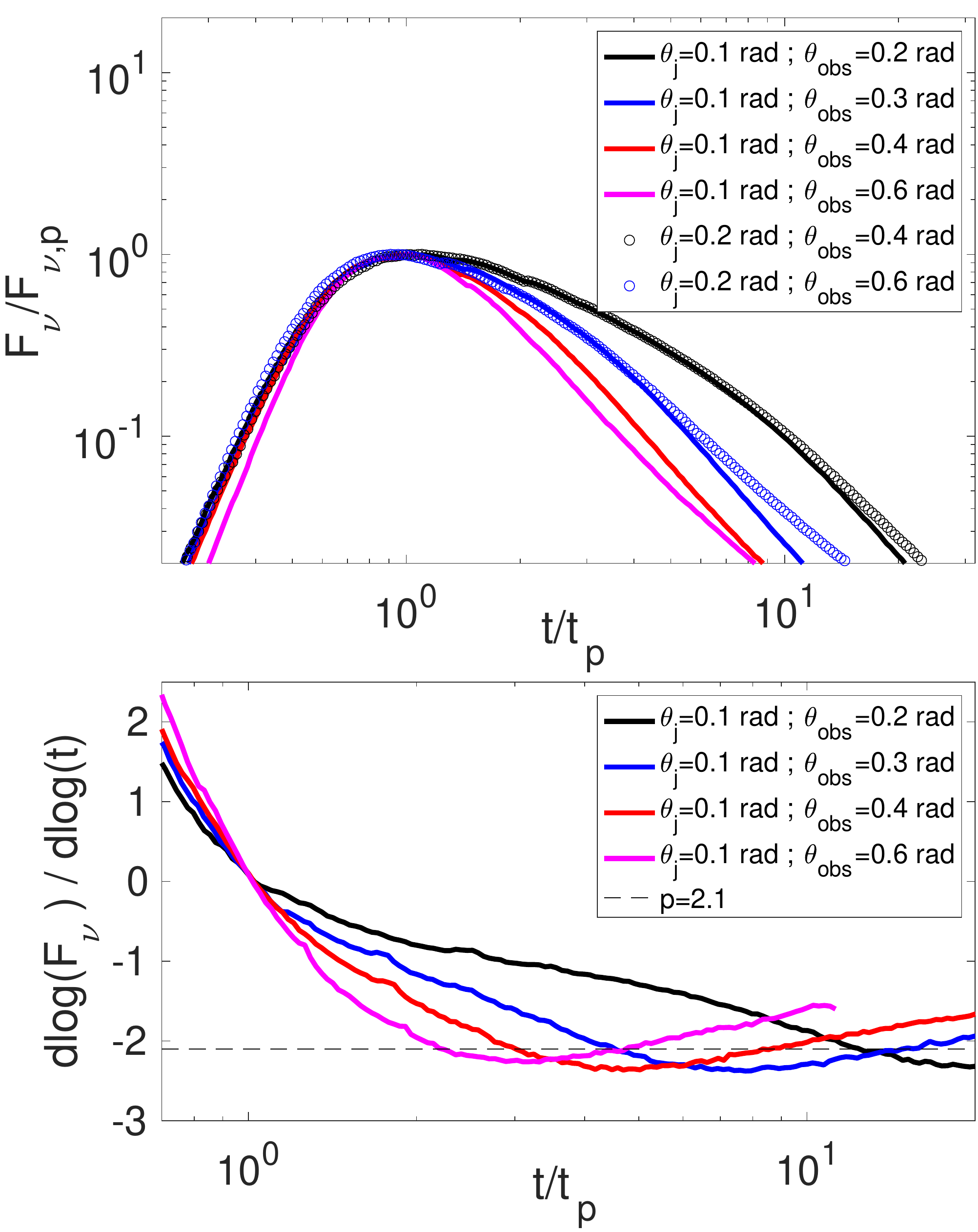}
\vskip -0.4cm
\caption{{ Top-hat jets with different jet opening angles and different ratios of viewing angle to jet opening angle. The light curves are calculated using the semi-analytic code described in \citet{Soderberg2006}. The jet is spreading sideways at the speed of sound in its rest-frame. {\it Top:} Light curves. Two jets with different $\theta_j$ and $\theta_{\rm obs}$ but the same $\theta_j/\theta_{\rm obs}$ 
that are depicted in the same colors (a solid line and a circled line)
have similar light curves as long as the emitting region is relativistic. 
Curves in different colors show that the shape of the peak depends on the ratio
$\theta_j/\theta_{\rm obs}$. For each color this ratio takes a different  value where the peak width becomes wider with increasing value of  $\theta_j/\theta_{\rm obs}$.
{\it Bottom:} The light curve slop (logarithmic derivative). The dashed horizontal line marks the asymptotic value, $p=-2.1$.  The intersections of the different light curves with this line marks the beginning of the asymptotic decline. Note that the slope drops slightly below the asymptotic value due to an overshoot that arises from the shape of the afterglow image (see discussion in \citealt{Granot2007}). The increase in the slope at late times is due to the transition to the sub-relativistic phase.}}
\vskip -0.5cm
\label{fig:width}
\end{figure}

{ To demonstrate the degeneracy, 
Figure \ref{fig:Gaussian} depicts a set of light curves with}
different values of $\theta_j$ and $\theta_{\rm obs}$ such that their ratio, $\theta_{\rm obs}/\theta_j=5$, remains constant. The light curves are identical as long as the emitting region is relativistic. 
In these curves $E_0$, the electron power-law index, $p$,  and the electron equipartition parameter  $\epsilon_e$ are kept constant. 
We scale the density, $n$, such that the peak time, $t_p$,  falls at the correct time, namely  $\Gamma(t)$ of the emitting region satisfies  $\Gamma(\theta_{\rm obs}-\theta_j) \approx 1$ around the peak. We also scale $\epsilon_B$, the magnetic equipartition parameter,  to obtain the correct peak flux $F_{\nu,p}$. 
While the  angles vary by a factor of 5,  $n$ and $\epsilon_B$ vary by five and three orders of magnitude, respectively.   
The deviation between the different curves at late time indicates the beginning of the transition to the sub-relativistic regime (see \S\ref{sec:sub-rel} below).  
{ The results shown in Figure \ref{fig:Gaussian} are for Gaussian jets but the degeneracy depicted is generic. It is independent of the jet's structure (see e.g. Figure \ref{fig:width}), the details of modeling
(full numerical simulations, semi-analytic modeling including or excluding sideways expansion etc.) and the exact frequency range in which the light curve is calculated, as long as it is within the same spectral regime.} 

{ Figure \ref{fig:width} shows the same degeneracy  for top-hat jets, where the afterglows are calculated using a model that includes the expected lateral expansion of the jet (at the speed of sound in the comoving frame; e.g., \citealt{Granot&Piran12}). 
Also here we see that as long as we keep the ratio $\theta_{\rm obs}/\theta_j = Const.$ the light curve remains unchanged. } 
These examples demonstrate one of the main points  of the paper. {\it A given light curve can be fitted with very different viewing and jet opening angles.}  All that is needed is to scale all angles and the Lorentz factor of the emitting region together. 
In the following section we explain the origin of this degeneracy.

\section{The General case} 
\label{sec:analytic}

In the previous section we have shown two specific examples of structured jets that are characterised by an angular structure, $E_{iso}(\theta)$. In both cases the entire jet structure was scaled  such that $E_{iso}(\theta) \rightarrow E_{iso}(f \theta)$ where $f$ is a constant. We have seen that in such a case when 
$\theta_{\rm obs} \rightarrow f\theta_{\rm obs}$ then by a proper choice of $n$ and $\epsilon_B$ the light curve remains unchanged. 

The origin of this degeneracy can be understood by following the contribution from each point along the shock. Consider a section of the shock that propagates radially at a Lorentz factor $\Gamma$ at a polar angle $\theta$ and azimuthal angle $\phi$ with respect to the jet axis. The angle between the shock velocity of this section and the line of sight (which is set at $\theta_{\rm obs}>\theta$ and  $\phi=0$) is $\psi(\theta,\phi, \theta_{\rm obs})$. The contribution of the emission from each section of the shock to the observed light curve depends on two factors, the rest-frame luminosity and the Lorentz boost to the observer. In the standard afterglow model, the former decreases with the observed time monotonically as a  power-law with a single break, as long as the emission at the observed frequency remains in a single power-law segment.  The break takes place when $\Gamma \psi \sim 1$.
The Lorentz boost also evolves as a broken power-law. It  rises sharply as a power-law at early times, as long as  $\Gamma \psi \ll 1$, and decreases as a power-law at late times when $\Gamma \psi \gg 1$. As a result, the observed emission from each section of the shock is a broken power-law that rises until $\Gamma \psi \approx 1$ and decays afterwards (note that for each section $\Gamma \psi \approx 1$ at a different observer time). The total observed light curve is obtained by integrating the contribution from  different sections of the jet. 
Now, let's rescale the jet according to $E_{iso}(\theta) \rightarrow E_{iso}(f \theta)$. With this mapping each section of the shock is mapped according to $\theta \rightarrow f\theta$, $\phi \rightarrow \phi$. If we also move the observer to  $\theta_{\rm obs} \rightarrow f\theta_{\rm obs}$ then for each point along the shock $\psi \rightarrow f\psi$. At the same time we rescale the Lorentz factor of the shock at each point, e.g., by changing $n$, such that $\Gamma(t) \rightarrow f^{-1}\Gamma(t)$, where $t$ is observer time. This mapping implies that for each section of the shock the rescaling conserves the evolution of $\psi \Gamma(t)$. Thus the evolution of the observed emission from each section (the rising and then decaying broken power-law) also remains the same. The rescaling of the jet energy ($E_{iso}(\theta) \rightarrow E_{iso}(f \theta)$) implies that the relative contribution of the different sections at any given time is also conserved. Thus, the light curve shape, which is obtained by integrating over the emission from all the sections remains the same as before the rescaling. The only thing which is changed is the overall normalization of the light curve. This is fixed by a rescaling of the microphysical equipartition parameters.   

As we saw, the degeneracy between the unknown afterglow parameters ($E$, $n$, $\epsilon_B$ and $\epsilon_e$) and the system geometry ($\theta_j$ and $\theta_{\rm obs}$), is present for any structured jet model{, which is defined by the angular structure of $E_{iso}(\theta)$,} where  the angular structure of the blast wave and the observing angle are assumed to go through a similar scaling. However, in reality, it is most likely that jets with different opening angles won't be  a simple rescaling of each other. The  structure of a jet 
is dictated by its evolution, depending on its properties at the injection site as well as its interaction with the merger sub-relativistic ejecta. Thus, it is reasonable to expect that jets with different opening angles also have a different overall structure which breaks the degeneracy discussed above. { Moreover, above we considered jets with structure in which there is only an angular dependency on the energy, i.e. $E_{iso}=E_{iso}(\theta)$. In reality, it is also possible that the initial Lorentz factor of the jet has an angular structure and in addition the jet may have a radial structure. If these additional components of the jet affect the light curve, then it is possible that the light curve has no degeneracy of the viewing angle and the jet opening angle.}
Nevertheless, since we don't have an a priori knowledge of the jet structure, even if the light curve is not degenerated, the light curve alone is not enough to determine $\theta_{\rm obs}$ and $\theta_j$ separately and only the ratio $\theta_j/\theta_{\rm obs}$ can be inferred. We show that by discussing the main observables and the information that each of those observables carries. We restrict our discussion to cases that are similar to GW170817. Namely jets with an angular structure that points away from the observer (i.e., $\theta_{\rm obs}>\theta_j$) where the emission at and following the peak is dominated by the jet core. We also assume that the afterglow is seen at a single power-law segment where $\nu_a,\nu_m<\nu_{\rm obs}<\nu_c$, where $\nu_a$, $\nu_m$ and $\nu_c$ are the characteristic synchrotron frequencies \citep[e.g.,][]{Sari+99}.  We note that 
There is one constraint, however,  that
can be derived: $\theta_{\rm obs} \lesssim 1$.  Otherwise the emission region during and after the peak won't be relativistic and the whole light curve structure will be different. 

\subsection{Peak time and flux} 
The most prominent observables are  the time and flux of the peak of the light curve.  \citet{Nakar2002} and \citet{Totani2002} have derived analytic relations for the peak flux and time for off-axis afterglows of top-hat jets. Remarkably, it turns out that these results are valid  for a wide range of jets, regardless of their angular structure, as long as the peak emission is dominated by the jet core \citep{Gottlieb2019}.

The light curve in such jets peaks when the jet core decelerates enough so that its relativistically  beamed  emission cone starts including the observer.  This happens when the Lorentz factor of the jet core satisfies $\Gamma (\theta_{\rm obs}-\theta_j) \approx 1$. The Lorentz factor of the core depends on the ratio $E/n$, where $E$ is the total core energy and $n$ the external density is assumed to be uniform. The peak time satisfies
\begin{equation}\label{eq:peak_timen}
t_{p} \propto \bigg(\frac{E}{n}\bigg)^{1/3} \bigg(\theta_{\rm obs}-\theta_j\bigg)^2
\ .
\end{equation}
We expect $E/n$ to vary by many order of magnitudes between different mergers and therefore, unless there is an additional constraint on $E/n$, the peak time  doesn't provide a measurement of  
$(\theta_{\rm obs}-\theta_j)$.

One can expect that the peak flux  provides additional information. However, the peak flux, $F_{\nu,p}$ depends on additional unknown microphysical parameters,
\begin{equation}
    F_{\nu,p} \propto E ~n^\frac{p+1}{4} \epsilon_e^{p-1} \epsilon_B^\frac{p+1}{4} \theta_{\rm obs}^{-2p} \nu^{-\frac{p-1}{2}} D^{-2} \, ,
    \label{eq:peak.mag} 
\end{equation}
where $ D $ is the distance to Earth and $ \nu $ is the observed frequency. $\epsilon_B$ 
is poorly constrained and may vary by several order of magnitude between different models. This freedom implies that $F_{\nu,p}$ does not provide a  useful constraint on the viewing angle as well.

\subsection{The rising light curve}
\label{sec:rising}

Numerous attempts to fit the light curve of GW170817  reveal that the light curve during the rising  phase depends on the structure of the jet at large angles, $\theta>\theta_j$ \citep{Nakar2018a,Alexander+18,Margutti+18,Dobie+18,DAvanzo+2018,Troja+18,Mooley+18a,Mooley+18b,Mooley+18c,
Granot2018,Lamb+2019,Wu2019,Fong+19,Ghirlanda+19,Troja+19,Hajela+19}. 
 This is true regardless of the origin of this structure, whether it arises due to the jet launching mechanism \citep[e.g.,][]{Kathirgamaraju2019,Fernandez2019,Christie2019,Nathanail2020}, the interaction of the jet with the ejecta  (the jet-cocoon; \citealt{Nakar2017,Lazzati2017a,Margutti+18}) or some other prameterized model of angular structure (e.g., Guassian or power-law).   At early times the afterglow is dominated by emission from a region at $\theta > \theta_j$ and only as the jet slows down its core begins to dominate the observed emission. Thus, the observed light curve during the rising phase is determined by the viewing angle as well as by the angular structure at $\theta> \theta_j$ \citep[e.g.,][]{Lamb2017}. As the latter is unknown it is impossible to determine the viewing angle from the light curve during this phase. 
 
Several  authors \citep[e.g.,][]{Ryan2020,Takahashi2020,Beniamini2020} examined recently  the information that can be drawn from this early phase on the jet structure. The main focus  of these papers  is the light curve phase that is dominated by emission from the ``wings'' of the jet,  $\theta> \theta_j$, that is beamed toward the observer (i.e., the emitting region is within an angle $1/\Gamma$ from the line of sight). \cite{Ryan2020} refer to this phase in the more general term ``structured phase'' since the flux during this phase may also decline slowly and not only rise. 
 
\cite{Takahashi2020} develop a method to invert  the light curve, finding a jet structure that can generate it. The key point  is that at any given observer time, $t$, the observed light curve is dominated by emission from   $\theta>\Theta(t)$, where $\Theta$  decreases monotonically with time. Thus, for a given viewing angle  $\theta_{\rm obs}$, various parameters such as $n$, $\epsilon_e$ and $\epsilon_B$,  and a known jet structure  at $\theta>\Theta(t_1)$
one can invert the  light curve  between $t_1$ and $t_2$  and determine  the structure at $\Theta(t_2)<\theta<\Theta(t_1)$.   \cite{Takahashi2020} show that  for every  $\theta_{\rm obs}$ there is a solution for the jet structure that is consistent with the observations. Moreover, since the only constraints on the jet structure at $\theta>\Theta(t_1)$ are derived from the first data point at $t_1$, there is an infinite number of possible structures at $\theta>\Theta(t_1)$ that are consistent with this data point. Therefor, for every viewing angle there is an infinite number of possible different jet structures that are compatible with the observed structured phase of the afterglow.

{ \cite{Ryan2020} took a different approach. They derived a closure relation between the temporal evolution of different spectral power-law segments. They find that for each spectral segment the temporal logarithmic derivative of the light curve is a function of a single parameter $g$, where $g$ is the same for all segments. For example, they find that ${d \log(F_\nu)}/{d\log(t)}={(3-6p+3g)}/{(8+g)}$ in  the segment $\nu_a,\nu_m<\nu_{\rm obs}<\nu_c$ which is most relevant for the observations of GW 170817. 
For  GW170817, during the structured phase ${d \log(F_\nu)}{d\log(t)}\approx 0.9$, and  $g \approx 8.2$. The factor  $g$ is determined by the jet structure and the observing angle. If we consider a jet structure parametrized by $N$ parameters\footnote{These parameters don't include the overall normalization of $E(\theta)$.}, $E=E(\theta,i_1,i_2,...,i_N)$, then a  rising afterglow phase that can be approximated by a single power-law with an effective value $g_{\rm eff}$ provides a single constraint on all these parameters and the viewing angle, $g_{\rm eff}=g_{\rm eff}(\theta_{\rm obs}, i_1,i_2,...,i_N)$. As \citet{Ryan2020} show,} {\it every} jet structure that satisfies the observed value of $g_{\rm eff}$ is consistent with the observations and therefore without having an a-priory knowledge of the functional form of the jet structure this rising (structured) phase does not provide any information on $\theta_{\rm obs}$, $\theta_j$ or on the ratio $\theta_{\rm obs}/\theta_j$.
 
There are interesting implications of this result for the two most commonly used jet structure parametrizations:  Gaussian and power-law jets. A Gaussian jet depends only on a single parameter, $\theta_j$ ($g$ is always independent of the overall normalization). Thus, $g_{\rm eff}$ is a function of two parameters, $\theta_{\rm obs}$ and $\theta_j$. \cite{Ryan2020} show that for a Gaussian jet $g_{\rm eff}={\theta_{\rm obs}^2}/{4\theta_j^2}$.
Thus, while fitting a Gaussian jet to the structured phase light curve does not provide information on $\theta_{\rm obs}$ and $\theta_j$ separately,  it  does measure $\theta_{\rm obs}/\theta_j$. On the other hand, a  power-law jet depends on two  parameters the core angle,
$\theta_j$, and the power-law $b$.  In this case  $g_{\rm eff}$ is a function  $\theta_{\rm obs}$, $\theta_j$ and $b$, and more specifically  $g_{\rm eff}=g_{\rm eff}({\theta_{\rm obs}}/{\theta_j},b)$. 
Thus, for a power-law jet the rising phase of the light curve does not determine $\theta_{\rm obs}/{\theta_j}$.

{ \cite{Takahashi2020} and \cite{Ryan2020} considered only jets with an angular structure $E_{iso}$, finding that in such jets that rising phase does not constrain the viewing angle and the jet opening angle. In general there is even more freedom in the jet structure. First it is possible that also the initial Lorentz factor depends on the angle. Second, at each angle there may be a radial structure. Namely, it is possible that there is a distribution of velocities and not only a single characteristic Lorentz factor. In the context of our discussion, it is clear that adding more freedom to the jet structure does not add additional information.} Thus, we conclude this sub-section with the understanding that that the rising phase of the light curve provide only loose constraints on the jet structure. There is a range of possible structures that can fit any given light curve and, unless there is an a-priory information on the jet structure, this phase provide no constraints on $\theta_{\rm obs}$, $\theta_j$ or the ratio between them.   
 
\subsection{The declining phase}
At and after the peak,  the light curve is dominated by the jet's core. The peak is followed by a rapid decline that reaches its asymptotic value once the jet decelerates enough so the beam of the emission from the entire jet core includes the observer. At that point the light curve is similar to the one seen by an observer along the jet axis
\citep{Sari+99}. Thus, during the decline the light curve is independent of the jet opening angle and of the viewing angle and therefore it contains no information about either one of those angles.

\subsection{The peak width} 
A more subtle light curve observable is the shape of the peak.   This is   a ``second order" observable as it can be estimated only for  bright enough events. The exact shape of the light curve near the peak depends on the unknown detailed structure of the jet near the jet core (at $\theta \approx \theta_j$) and therefore there is some
freedom in modeling it. However,  one observed feature, the width of the peak, does provide information about the ratio $\theta_j/\theta_{\rm obs}$.  
More precisely, the time between the peak itself and the  moment when  the decay becomes comparable to the one seen by an observer along the jet axis. We denote this time as $\Delta t_p$.  

The top panel of Figure \ref{fig:width} shows several light curves of top-hat jets with different opening angles and/or viewing angles. It shows, first, that for a given jet opening angle a larger viewing angle results in a narrower peak, i.e., smaller $\Delta t_p/t_p$. Second, it shows that afterglows with different opening angles and different viewing angles produce similar curves if they have the same $\theta_j/\theta_{\rm obs}$ ratio. Thus, the width of the peak depends on a single parameter, the ratio $\theta_j/\theta_{\rm obs}$. 

To understand this behaviour recall that the peak is observed roughly at the time where $\Gamma(t_p) (\theta_{\rm obs} - \theta_j) \approx 1$. The light curve decay becomes similar to the one seen by an observer that is along the jet axis when  the beaming is such that the observer at $\theta_{\rm obs}$ detects radiation from the whole core of the jet, namely $\Gamma(t+\Delta t_p) (\theta_{\rm obs} + \theta_j) \approx 1$. For an on-axis observer the Lorentz factor of the emitting region decays as a power-law in the observer time, $\Gamma(t) \propto t^{-k}$ ($k=1/2$ for a spreading jet  and $k=3/8$ if  spreading is ignored, see e.g. \citealt{Granot2007}). This implies that the time between the peak and the asymptotic decay satisfies
\begin{equation} 
 \frac{t_p + \Delta t_p }{t_p}  \approx  \bigg [\frac{\Gamma(t+\Delta t_p)}{\Gamma(t_p)} \bigg]^{-1/k} \approx \bigg [\frac{1+\frac{\theta_j}{\theta_{\rm obs}}}{1-\frac{\theta_j}{\theta_{\rm obs}}} \bigg]^{1/k} \, . 
 \label{eq:width}
\end{equation}

Equation \ref{eq:width} explains the dependence of $\Delta t_p$ on the ratio $\theta_j/\theta_{\rm obs}$ and why it depends only on this factor and not on $\theta_j$ or $\theta_{\rm obs}$ alone. It also quantifies this dependence. The bottom panel of Figure \ref{fig:width} depicts the evolution of light curve slope (logarithmic derivative) from several top-hat jets, which are based on light curve modeling by jets that are free to spread sideways (corresponding to  $k=1/2$ in Eq. \ref{eq:width}).
The increase of the slope seen at late times is due to the beginning of the transition to the sub-relativistic phase. The agreement with Eq. \ref{eq:width} is good.  This equation predicts $(t_p + \Delta t_p)/t_p$=[1.96, 2.8, 4 and 9] for $\theta_{\rm obs}/\theta_j=[6, 4, 3, 2]$. The numerical calculations shown in this figure find that the asymptotic slope is reached at comparable values. 

Equation \ref{eq:width} is applicable not only to top-hat jets. It provides a good  approximation to any jet with an angular structure whose peak is dominated by the jet core, such as in GW170817. The reason is that the emission at and following the peak, in such jets,  is similar to the one seen from a top-hat jet. Note, however, that the approximation is accurate only to within a factor of order of unity, that vary between different jet structures. First the the shape width of the peak does vary somewhat between different jet structures (and different accuracy of the hydrodynamic modeling). Second, the exact definition of $\theta_j$ vary between different jet models.

\subsection{Polarization}
A different ``second order" observable is the polarization. Like the width of the peak,   polarization measurements require a relatively strong signal and it won't be detected in most cases. Even in GW170817  only an upper limit on the radio polarization  was found \citep{Corsi2018}. A synchrotron emitting jet can produce linearly polarized emission \citep{Sari1999,Ghisellini1999}. For an observer at $\theta_{\rm obs} > \theta_j$, the polarization peaks when the jet's core becomes visible, namely around $t_p$. As discussed earlier we cannot determine the geometry from $t_p$. The magnitude of the polarization  depends on the unknown structure of the magnetic field within the emitting region.  Hence, like other measures of the afterglow light curve,   polarization observations  won't be useful to determine the overall  geometry. 

\subsection{Additional constraints on the afterglow parameters}\label{sec:AditionalConstraints}
In some mergers it may be possible to obtain additional constraints on the afterglow parameters. For example, it is possible that the observed afterglow spectrum will enable an identification of characteristic break frequencies such as $\nu_c$ or $\nu_m$. Another example is a constraint on $n$ that may be obtained by observations of the host galaxy at the location of the merger. How many such constraints are needed in order to determine the geometry? As we have shown above, the light  curve provides three constraints, Eqs. \ref{eq:peak_timen}-\ref{eq:width} while the model has six free parameters $E$,$n$,$\epsilon_B$, $\epsilon_e$, $\theta_j$ and $\theta_{\rm obs}$ ($p$ is measured from the spectrum). Therefore, in general, we need three additional constraints to find the geometry. Even if we fix the value of $\epsilon_e$, which is the  best constrained parameter, two additional constraints are needed. Thus, a single additional constraint on the afterglow parameters, such as an identification of a break frequency or an external constraint on $n$, won't be enough to measure  $\theta_j$ and $\theta_{\rm obs}$. 

\section{Breaking the degeneracy} 
\label{sec:breaking}
The degeneracy between the system geometry and afterglow parameters can be removed by  an independent measurement of the Lorentz factor. There are two ways to obtain this: measuring the centroid motion of the afterglow image using radio VLBI observations or identifying the late time light curve transition to the sub-relativistic phase. Both methods are ``second order" observables as they require a bright afterglow with observed peak flux that is significantly brighter, by at least an order of magnitude, than the detector threshold. One could expect that a bright burst implies necessarily a nearby one. However, there are two other   critical factors. 
The peak flux depends very strongly on the viewing angle (Eq. \ref{eq:peak.mag}). Therefore, a bright afterglow requires a relatively small viewing angle. For example, the afterglow of GW170817 would have been barely detectable if its viewing angle would have been larger by a factor of 1.5 \citep{Gottlieb2019}. A second possible factor that leads to a bright signal is a large circum-burst density (see Eq. \ref{eq:peak.mag}). 

\subsection{Centroid motion}
The image of a relativistic jet that is seen at some angle with respect to the line of sight can move  at an apparent superluminal velocity $\beta_{\rm app} > 1$ \citep{Rees1966}. This phenomena was observed first in AGNs in the 60ies and was first discussed in the context of GRBs by \cite{Sari1999}. This superluminal motion enables us to estimate the Lorentz factor of the jet. The best, and also easiest, way to measure $\Gamma$ is to obtain two VLBI images around the time of light curve peak. At this time the image is at an angle $\theta_{\rm obs}-\theta_j \approx 1/\Gamma$ with respect to the line of sight and the apparent velocity of the image is $ \beta_{\rm app} \approx \Gamma$ \citep{Mooley+18b}. 

\subsection{The sub-relativistic regime} 
\label{sec:sub-rel}
An inspection of Figure \ref{fig:Gaussian} reveals that the light curves deviate at late time. This deviation occurs because different jets reach the sub-relativistic regime at different times. Once $\Gamma\approx 1$ the dynamics of the blast wave changes and the beaming becomes unimportant, leading to a more moderate decay rate, thereby breaking the degeneracy mentioned earlier. Thus, observations of the light curve sufficiently late, so that this regime is seen, can reveal the earlier values of $\Gamma$ and with them we can determine the geometry. 

Unfortunately, like the centroid motion, the late time light curve is also a ``second order" observable that can be detected only for very bright  events. Even for GW170817 it was not detected.  Although there may be a hint for this transition in the latest X-ray observations \citep{Hajela2020}, they are so faint  that  the uncertainties are too large to identify or exclude its presence.

\section{GW170817} 
\label{sec:GW170817}

Next we discuss the measurements of $\theta_j$ and $\theta_{\rm obs}$ in GW170817 in view of the results presented above. The afterglow observations of GW170817 included a detailed radio to X-ray light curve (see \citealt{nakar2019} for a review and references therein). The emission in all these bands was on a single power-law segment which corresponds to an electron power-law index $p \approx 2.15$. The light curve showed a steady continuous rise with a rising slope $\alpha_r \approx 0.8$ that peaked at $t_p \approx 130-150$ days and was followed by a rapid power-law decay that reached a steady slope $\alpha_d \approx -p$, roughly within $\Delta t_p \approx 100-150$ days.
VLBI observations at days 75 and 230, just before and just after the peak, have shown a clear motion of the radio image centroid at an apparent velocity $\beta = 4.1 \pm 0.5$ \citep{Mooley+18b}, a result that was later confirmed by additional VLBI image taken on day 207 \citep{Ghirlanda+19}.

\begin{figure}
\centering
\includegraphics[scale=1.]{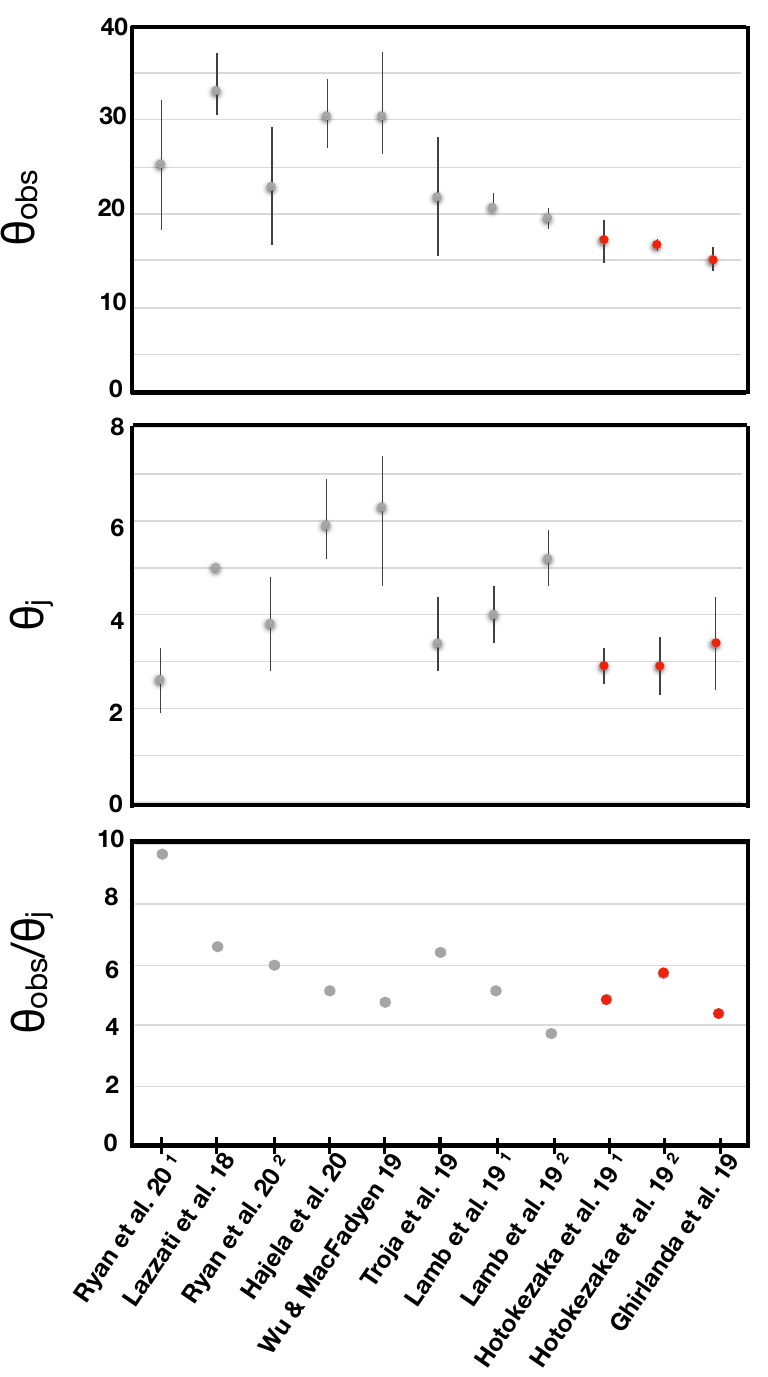}
\vskip -0.2cm
\caption{Results of recent modeling of the geometry of GW170817, {\it Top:} viewing angles  {\it Middle:} jet opening angles  {\it Bottom:} their ratios. Error bars are from the original papers. Error bars are not shown on the ratio $\theta_{\rm obs}/\theta_j$ as the 
quoted errors in the original papers are most likely correlated. 
While values for $\theta_{\rm obs}$ or $\theta_j$ vary strongly, the  scatter in $\theta_{\rm obs}/\theta_j$ is 
small with almost all values are around $\theta_j \approx 5-6^o$.   
The last three models (marked in red)  \citep{Hotokezaka2019,Ghirlanda+19} use the VLBI information. Some papers give two values and those are marked by superscripts $^{1,2}$. }
\vskip -0.5cm
\label{fig:fitting}
\end{figure}

These observations were modeled by numerous authors, using various modeling techniques. The majority of these papers modeled only the light curve, ignoring the VLBI data, while a  few  modeled both the light curve and the VLBI data. Figure \ref{fig:fitting} depicts the constraints that these papers derived on $\theta_{\rm obs}$ and $\theta_j$. It shows a large scatter between the different measurements. Moreover, the various measurements are often inconsistent with each other. This is not surprising as  Figure \ref{fig:Gaussian} already demonstrated that the light curve can be well fitted with very different angles.  Figure \ref{fig:fitting} also shows  the measurements of $\theta_{\rm obs}/\theta_j$, finding a much better agreement between the various papers as most measurements find $\theta_{\rm obs}/\theta_j \approx 5-6$. 

A comparison of the results on GW170817 to our findings brings up two points, one showing an agreement and the other a disagreement. First, the value of $\theta_{\rm obs}/\theta_j$ found by most of the studies is in remarkable agreement with Eq. \ref{eq:width}. In GW170817 $(t_p+\Delta t_p)/t_p \approx 2$, which corresponds to $\theta_{\rm obs}/\theta_j \approx 5.8$ for $k=1/2$. Second, there is a disagreement between our finding here that the light curve alone cannot constrain the values of $\theta_{\rm obs}$ and $\theta_j$, and the fact that many papers do exactly that. The natural question is how these measurements are derived and what is the origin of the disagreement. When each of the papers that measures $\theta_{\rm obs}$ and $\theta_j$ is examined it is found that the constraints on the geometry have different sources, but in most cases they boil down to the assumed arbitrary priors of the models. { In some cases  the modeling includes additional measurements that are unrelated to the light curve, such as VLBI or GW data.  In the case of GW170817, apart for a VLBI superluminal motion, there is no information that can be used to tightly constrain either $\theta_{\rm obs}$ or $\theta_j$.}
To demonstrate that, we discuss, as  examples, the results of  four specific papers,
\cite{Lazzati+2018,Troja+19}, \cite{Hajela2020} and \cite{Ryan2020}.

\cite{Lazzati+2018} used a combination of a hydrodynamic simulation and  semi-analytic modeling to calculate the  light curve. The simulation is used to find the jet structure as it emerges from the sub-relativistic ejecta and the semi-analytic model follows  the blast wave that the jet drives into the circum-merger medium and the emission that it produces. They conclude that the observer is at an angle of $33^{o~+4}_{-2.5}$. The origin of this measurement is that these authors implicitly assumes  the value of $\theta_j$. The fit is based on a single hydrodynamic simulation whose initial conditions produce a jet with an opening angle $\theta_j=5^o$.  Having a fixed value for $\theta_j$ the value of $\theta_{\rm obs}$ is then determined by the light curve constraint on $\theta_{\rm obs}/\theta_j$.

\cite{Troja+19} run a large set of semi-analytic models of Gaussian jets. 
They include in their modeling 
a lower limit on $\nu_c$ based on the X-ray data and
used the GW inclination measurement to mitigate the degeneracy between the angles. They find  a viewing angle $\theta_{\rm obs} = 22^o \pm 6$  and a jet core opening angle $\theta_j=3.4^o \pm 1.1$. When examining the MCMC used to derive these measurements it is found that, as expected, the only contribution of the light curve is by determining the ratio $\theta_j/\theta_{\rm obs}$. This can be seen from their “corner plot” that shows all two-dimensional projections of the posterior probability density functions obtained by the MCMC simulation (see their Fig. S2, in supplementary information). 
The two-dimensional projection of $\theta_j$ and $\theta_{\rm obs}$ shows a very tight linear correlation where $\theta_{\rm obs}/\theta_j \approx 6$. Similar correlation was already identified by \cite{Troja+18}, which carried out a similar analysis. The low probability of $\theta_{\rm obs}$ at small values  is dominated by the prior chosen, $P(\theta_{\rm obs}) \propto \sin(\theta_{\rm obs})$, while the low probability of $\theta_{\rm obs}$ at large values is a result of the GW constraints.

A recent paper that constrain the geometry of GW170817 is \cite{Hajela+19}. This paper uses in addition to the light curve also a lower limit on $\nu_c$, based on the X-ray data, and an upper limit on $n$ based on observation of the hot gas in the host galaxy at the location of the merger. The measurements that they obtain are $\theta_{\rm obs} = 30.4^{
o~+4}_{-3.4}$  and $\theta_j=6.66^{o~+2.48}_{-1.31}$ . The jet model that they use for the light curve fitting is "boosted fireball", which is  parameterized structured jet model with two-parameters  \citep{Duffell2013}. The details of this structured jet model are unimportant for this discussion, except for the point that one of the model parameters, $\gamma_B$, is roughly the inverse of the jet opening angle, namely $\theta_j \approx 1/\gamma_B$. The origin of the limits on the geometry can be seen by examining the MCMC corner plot of their model (their Fig. 4). It shows that the upper limit on $\theta_{\rm obs}$ is obtained by a combination of the light curve, the limits on $\nu_c$ and $n$ and the natural upper limit $\epsilon_e <1$. As discussed in \S\ref{sec:AditionalConstraints}, combination of the light curve with three additional observables  may provide a constraint on the geometry. The lower limits on $\theta_{\rm obs}$ and $\theta_j$ are, however, artificial and they are derived directly from the selected prior for $\gamma_B$ that (with no physical justification) has a maximal value of $\gamma_B=12$. This prior translates directly to $\theta_j>4.5^o$. An examination of the two dimensional projection of the posterior probability density functions of $\gamma_B$ and $\theta_{\rm obs}$ shows  a tight linear correlation between $\theta_{\rm obs}$ and $1/\gamma_B$, where $\theta_{\rm obs}\gamma_B \approx \theta_{\rm obs}/\theta_j \approx 5.5$. Thus the artificial prior  $\gamma_B<12$, together with the light curve, imposes   $\theta_{\rm obs} \gtrsim 25^o$.

Finally, \cite{Ryan2020} compare Gaussian and power-law jets to the observed light curve of GW170817. Their priors on $\theta_{\rm obs}$ are similar to those taken by \cite{Troja+19} and the constraints on $\theta_{\rm obs}$ result from these priors. \cite{Ryan2020} highlight the tight constraints that they obtain on $\theta_{\rm obs}/\theta_j$ in both jet models. However, the values they find for the two models are significantly different. For the Gaussian jet they find $\theta_{\rm obs}/\theta_j=6.12 \pm 0.18$. This value is similar to the constraint they obtained from $g_{\rm eff}$ based on the rising phase (see \S\ref{sec:rising}) and is also consistent with one 
derived from the  light curve peak (eq. \ref{eq:width}). Those two constraints are completely independent and the fact that both are giving similar values is a coincidence (unless the jet is really a Gaussian). Moreover, this is the reason that a jet structure with a shape that depends on a single parameter, such as the Gaussian jet, can fit the entire data. For the power-law jet 
they find $\theta_{\rm obs}/\theta_j=9.38^{+0.73}_{-0.56}$. Here the value of 
$g_{\rm eff}$ depends on two parameters $\theta_{\rm obs}/\theta_j$ and the power-law index $b$, where for a given $g_{\rm eff}$ the value for $\theta_{\rm obs}/\theta_j$ is smaller for larger values of $b$. When the entire light curve is considered the width of the peak determines $\theta_{\rm obs}/\theta_j$, which does not depend strongly on the jet functional structure (e.g., a Gaussian or a power-law). Thus, in case of a power-law jet the combination of $g_{\rm eff}$ and $\theta_{\rm obs}/\theta_j$
should determine $b$. The value of $\theta_{\rm obs}/\theta_j$ that \cite{Ryan2020} find for the power-law jet is larger than the values that they find for a Gaussian jet. The difference is only of order unity (a factor of 1.5), and in principle the constraints from the two models are not expected to be exactly the same. Nevertheless, it seems that the reason for the difference is, at least partially, the prior that they take for the power-law index $b=[0-10]$. As seen in their corner plot (their figure C2),  the posterior distribution of $b$ peaks at $b=10$, implying that there is a better fit then the one they find for a power-law jet with $b>10$ and $\theta_{\rm obs}/\theta_j<9$.

Three papers incorporated the VLBI data in their analysis \citep{Mooley+18b,Hotokezaka2019,Ghirlanda+19}. This data have shown that at the time of the peak $\Gamma \approx 3-5$, implying $\theta_{\rm obs}-\theta_j \approx 0.2-0.35 {\rm rad} \approx 12-20^o$. Together with the light curve constraint, $ \theta_{\rm obs}/\theta_j \approx 5-6$, one obtains $\theta_{\rm obs} \approx 14-24^o$  and $\theta_j \approx 2-4^o$.  \cite{Hotokezaka2019} and \cite{Ghirlanda+19} fitted the data with  semi-analytic models of Gaussian and power-law structured jets. They obtained for all the models $\theta_{\rm obs} \approx 16^o$  and $\theta_j \approx 3^o$ with a $1\sigma$ uncertainty of $\sim 5-10\%$. \citet{Mooley+18b} used general analytic considerations (largely independent of the jet structure) and a large set of numerical hydrodynamic simulations to find 
$\theta_{\rm obs}=14-28^o$  (where the most likely value is $\theta_{\rm obs} \approx 20^o$ ) and $\theta_j < 5^o$. This range of viewing angles includes all of their numerical jet models that are consistent with the VLBI data at $2\sigma$. The reason for the large difference in the uncertainties obtained by the two methods is that the error reported by  \cite{Hotokezaka2019} and \cite{Ghirlanda+19} is the statistical error of the specific models they explored. Namely, this is the error obtained by the fit to the data under the assumption that the jet model used is the correct one. This estimate does not include the systematic uncertainty arising from the fact that the jet structure may be (and almost certainly is) different than the specific structures they explored. \citet{Mooley+18b} find that this modeling error is the dominant source of uncertainty, and approximating the range of of viewing angles that they report as a $2\sigma$ interval, we estimate their $1\sigma$ error at $\sim 15$\%

\section{Implications to measurements of $H_0$}
\label{sec:Hubble}

The results discussed above have far reaching implications to a seemingly unrelated issue, the measurements of $H_0$ using GW and EM observations of BNS mergers. 
A precise determination of the  inclination angle of the binary can significantly contribute to the accuracy of this measurement.
Assuming that the jet is aligned with the binary orbital angular momentum the viewing angle $\theta_{\rm obs}$ equals to the binary inclination angle and hence its importance.  

The detection of GW170817 enabled the LVC team to estimate $H_0$ using the GW observations and the redshift information \citep{Abbott2017hubble} at a $1\sigma$ accuracy of $\sim 15\%$. 
Shortly afterwards, \cite{Hotokezaka2019} demonstrated the importance of a viewing angle determination improving the  accuracy of the $H_0$ measurement by a factor of 2 to a level of $\sim 7\%$. 
These $H_0$ measurements, were followed by various estimates of the future potential of this method to measure $H_0$ at accuracy that compete with other methods \citep[e.g][]{Chen2018,Sathyaprakash2019,Mortlock2019,Nicolaou2019}. 
Here we discuss the implications of 
our finding on EM estimates of the viewing angle  to future measurements of $H_0$.

The chirping  GW signal from a BNS merger enables us to measure directly $D_L$, the luminosity distance of the source, skipping various distance ladders used in other methods. The crux of the matter here is the dependence of $D_L$ on the viewing angle: 
\begin{equation}
   D_L \propto  \frac{\cos(\theta_{\rm obs})}{h },
   \label{eq:DL}
\end{equation}
where $h$ is the GW strain, and we have assumed as mentioned above, that the jet is aligned with the binary's axis.  This relation is approximate and it holds for  small viewing angles,
$\theta_{\rm obs} \ll 1$ rad. For $\theta_{\rm obs} \gtrsim 1$ rad,  $\cos(\theta_{\rm obs})$ is replaced by  more complicated geometric functions of $\theta_{\rm obs}$ that depends on the GW polarization. In principle the ratio of the GW strain of the different polarizations carries information on the viewing angel. However, 
this ratio $\approx 1$ for small angles and  it can be used  to measure $\theta_{\rm obs}$ only for  $\theta_{\rm obs}\gtrsim 70^o$  \citep{Chen2019}.

We have shown in \S \ref{sec:breaking} that determination of the viewing angle is possible only for bright bursts and in particular it is most effective for small viewing angles. This is convenient since first, this is exactly the range for which the GW measurements of the viewing angles are impossible. Moreover,  the error in the viewing angle, $\delta \theta_{\rm obs}$, is typically smaller for smaller viewing angles, and this error is translated to an error of $\delta \cos(\theta_{\rm obs}) \approx \theta_{\rm obs} \delta \theta_{\rm obs} $. For example, a 15\% error of the viewing angle measurement at $\theta_{\rm obs}=20^\circ~[30^\circ]$ is translated to an error of 2\% [5\%] in $\cos(\theta_{\rm obs})$. Thus, we can expect a very precise determination of $ \cos(\theta_{\rm obs})$ and as a result of $D_L$ at small angles (see Eq. \ref{eq:DL}). 
However, the need of a  small viewing angle 
has the obvious disadvantage that such events are relatively uncommon. This is compensated somewhat by the enhanced emission of GWs along the systems axis \citep[e.g.][]{Sathyaprakash2009} that increases the number of mergers observed at small angles. Still, 
only a small fraction of the BNS mergers with EM emission will have a bright enough afterglow to constrain the viewing angle.

Before turning to examine the implications of this result we should consider the other   
major sources of uncertainty in the measurement of $H_0$. 
As evident  from Eq. \ref{eq:DL} a second source of error, $\delta h$,  arises from  the GW strain measurement. $\delta h$  is roughly the inverse of the GW signal-to-noise ratio
\citep{Chen2019}. Clearly $\delta h/h$ increases with the distance and it will decrease with future improvement of GW detectors. 
A third source of error arises from contamination of the 
redshift measurement of the host by  its  peculiar velocity, $\delta v_p$.  The uncertainty in the peculiar velocity is roughly $150-250$ km/s and it is independent of the distance to the source. As the cosmological velocity increases with distance, the relative importance of $\delta v_p$  compared to the measured host receding velocity decreases with the distance. 

The dependence of these last two error sources on the distance implies that there must be a sweet spot at intermediate distance in which  both are comparable. The exact position of this spot depends on the sensitivity of the GW detector and on the sophistication in determination of the peculiar velocity \citep[see e.g.][]{Chen2019,Nicolaou2019}. 
However, these errors dominate only if the viewing angle error $\delta \cos(\theta_{\rm obs})$ is sufficiently small, namely if the viewing angle was estimated accurately using EM observations. Otherwise the viewing angle error will dominate. 

To understand the relative importance of the three error sources we consider $H_0$ measurements based on  GW 170817, the only observed GW-EM BNS merger so far.  Without the EM measurement of $\theta_{\rm obs} $ the error in $H_0$ was about 15\% (1$\sigma$) and it was dominated  by the viewing angle error. The EM limits on $\theta_{\rm obs}$ dropped  $\delta \cos(\theta_{\rm obs})$ to $\sim 2$\% and  the overall uncertainty in  $H_0$ dropped to about $7\%$.  The error now was  dominated by the host peculiar velocity $\delta v_p$ with a non-negligible contribution arising from $\delta h$ 
\citep{Hotokezaka2019}. 
Overall, the distance of GW170817 was close to optimal with the sensitivity of O2 run, where only a small improvement in the accuracy of $H_0$ could have been made if it was slightly farther away (thereby slightly reducing the peculiar velocity error and increasing the error in $h$).

In the future, the improved sensitivity of the GW detectors will  enable a more accurate measurement of $H_0$ in events with VLBI observations.
A merger at a distance of $\sim 80-100$ Mpc, which is observed at a small angle where the GW signal is the strongest,  can have a low errors in $h$ (about $2-3\%$) once LIGO-VIRGO will achieve the design sensitivity, and a comparable  peculiar velocity error.
This distance is also optimal for VLBI detection of events at a viewing angle of $\sim 20^\circ$ \citep{Dobie2019}. Thus, assuming an error in $\cos\theta_{\rm obs}$ that is similar to GW 170817, we expect for such events a total uncertainty in $H_0$ of  $\sim 4-5\%$.

To conclude we expect two basic types of $H_0$ measurements. 
The majority won't have a sufficiently bright  EM signal and their error will be dominated by the viewing angle error, $\delta \cos(\theta_{\rm obs})$ which would be of order $15\%$ on average \citep{Chen2019}. For a small fraction of the events, the viewing angle will be measured accurately and for those the error will be determined by a combination of the strain error $\delta h$ and the peculiar velocity error $\delta v_p$, and it can be as low as $\sim 4-5\%$

A single event with an error $\sigma$ is statistically equivalent\footnote{There are two caveats concerning these rare mergers where the viewing angle can be measured. First, small samples are more susceptible to sampling bias. Therefore, $n$ events with a higher error are in some sense better than an ``equivalent" single event with a  smaller (by a factor $1/\sqrt{n}$) error.  Second, the inclusion into the analysis of $\theta_{\rm obs}$ determined by EM estimates introduces  systematic errors that are associated with the afterglow modeling and were not present otherwise.} to  $n^2$ events with error $n\sigma$.  
A single high precision merger would equal  to $\sim 10$ weaker ones\footnote{But still sufficiently bright  to detect an EM signal that will enable  the host redshift measurement.  Note that it is clear  (as was evident by the LVC O3 run) that most future 
BNS mergers won't have EM observations and those that will won't have an EM signal at the same quality as GW170817.}, each one constraining $H_0$ at a precision of 15\%. It is possible that this small fraction of high precision mergers will contribute to the precision in which we can measure $H_0$ using gravitational waves.

\section{Conclusions} 
\label{sec:conclusions}
A full interpretation of BNS merger observations requires the determination of  geometry of the system, the viewing angle and to a lesser extent the jet opening angle. As such, it is not surprising that numerous efforts were made to determine these quantities for GW170817, the first merger observed. Remarkably, different estimates that are largely based on modeling of the afterglow light curve vary by a factor larger than two and are inconsistent with each other (in terms of the estimate errors). Previous papers \citep{Ryan2020,Takahashi2020} have shown that in Gaussian and power-law jets there is some level of degeneracy between the viewing angle and the jet opening angle. Here we show that the degeneracy is complete and general to all jet structures, so the light curve on its own cannot constrain any of the angles. The reason stems from the fundamental degeneracy of relativistic outflows between the system geometry and the outflow Lorentz factor. The estimates of the geometry that were based on the light curve alone were mostly based on the priors assumed during the modeling and not on the light curve data. We have shown here further, that using the light curve, and in particular the width of the peak (see Eq. \ref{eq:width}), one can determine the ratio between the viewing angle and the jet opening angle.

It is important to note that the degeneracy that we demonstrate here has implications not only the  geometry of the jet, that we focused on in this paper. Its implications apply to determination of all other  parameters of the system, like the energy, the equiparitition parameters and the external density. All these parameters cannot be uniquely determined form the light-curve alone and additional information is essential (see also \citealt{Beniamini2020, Ryan2020, Takahashi2020}).

Measurements of the geometry require additional information.
This can be provided by observations of superluminal motion of the image centroid or the late time transition into the sub-relativistic regime. The first depends on VLBI observations, the second on detection of the light curve at very late time. Both require a stronger afterglow signal, which is brighter at the peak by at least an order of magnitude than the detection threshold. Thus, these measurements will be possible only for relatively rare events. 

Independent determination of $H_0$ is among the  most interesting  implication of BNS merger observations. A major source of uncertainty in this measurement is the binary viewing angle and therefore an estimate of this angle from the EM emission can significantly improve its accuracy. Our results imply that an EM determination of this viewing angle will be possible only for rare bright events. On the other hand the resulting error in $H_0$ using the additional information can be significantly lower (by about a factor of 3-4) than the one obtained using the GW signal and the redshift alone.

\section*{Acknowledgements}
We thank Chi-Ho Chan, Ofer Lahav, Kevin Lamb, Shiho Kobayashi, Geoffrey Ryan and Elenora Troja for helpful comments. This research is partially supported by  an advanced ERC grant TReX (TP) and  by a consolidator ERC grant JetNS  and an ISF grant (EN).




\bibliographystyle{mnras}

\end{document}